\documentclass[twocolumn,prl,aps,superbib,tightenlines,floatfix]{revtex4}
\usepackage{amsfonts}
\usepackage{amsmath}
\usepackage{amssymb}
\usepackage{graphicx}
\begin{document}
\topmargin -24mm
\textheight 250mm
\bibliographystyle{apsrev}
\title{Shot Noise in Nanoscale Conductors From First Principles}
\author{Yu-Chang Chen and Massimiliano Di Ventra\cite{MD}}
\affiliation{Department of Physics, Virginia Polytechnic Institute and State University,
Blacksburg, Virginia 24061}
\begin{abstract}
We describe a field-theoretic approach to calculate quantum shot noise in
nanoscale conductors from first principles. Our starting point is the
second-quantization field operator to calculate shot noise in terms of
single quasi-particle wavefunctions obtained self-consistently within
density functional theory. The approach is valid in both linear and
nonlinear response and is particularly suitable in studying shot noise in 
atomic-scale conductors. As an example we study shot noise in Si
atomic wires between metal electrodes. We find that shot noise is 
strongly nonlinear as a function of bias and it 
is enhanced for one- and two-Si wires due to the large 
contribution from the metal electrodes. For longer wires 
it shows an oscillatory behavior for even and odd number of atoms with opposite trend 
with respect to the conductance, indicating that current fluctuations persist with 
increasing wire length. 
\end{abstract}
\maketitle
The subject of steady-state current fluctuations (shot noise) has attracted
much attention from both theorists and experimentalists over the past
decade~\cite{buttiker}. Shot noise is due to charge quantization, and, as
such, is generally unavoidable even at zero temperature. While in signal
processing shot noise might be considered as an undesirable effect that blurs
the signal detection, in mesoscopic and nanoscale systems it is a physical
mechanism that can provide useful information to probe the electron energy
distribution~\cite{bulashenko}, the kinetics of electrons,~\cite{landauer} and
electron interactions due to the Coulomb repulsion and the Pauli exclusion
principle~\cite{liu} .

When electrons in a conductor diffuse in a completely uncorrelated way, the
shot noise magnitude reaches the well known Poisson limit $2eI$, where $e$ is
the electron charge and $I$ is the dc current.~\cite{schottky} In all other
cases, shot noise is proportional to the average current times a real number
$F$, called Fano factor.~\cite{buttiker} Except for some special cases, like,
e.g., transport in resonant-tunneling diodes, where shot noise is
significantly enhanced in the negative differential resistance region, due to
enhanced electron interaction in the well~\cite{pellegrini}, the Fano factor
is generally lower than
one.~\cite{ting,martinis,pardo,beenakker,nagaeve,nagaeve2,
loss,mishchenko} In a two-terminal conductor with $n$ channels, the Fano
factor can be expressed in linear response and at zero temperature in terms of
the transmission coefficients $T_{n}$ of each channel $n$%
~\cite{lesovik,buttiker2,martin2}
\begin{equation}
F=\sum_{n}T_{n}\left(  1-T_{n}\right)  /\sum_{n}T_{n},\label{eqbutt}%
\end{equation}
where the denominator is simply proportional to the average current. In
principle, if the transmission probabilities $T_{n}$'s are known, shot noise
can be evaluated using Eq.~(\ref{eqbutt}). However, the transmission
probabilities, and consequently shot noise, are an interrelated function of
the electronic and ionic distributions which are generally nonlinear in the
external bias.~\cite{mioprl,guo,bula} These distributions need to be
calculated self-consistently at all external voltages, and can not be simply
derived from ground state calculations.~\cite{mioprl,bula} The last point is
particularly relevant in nanoscale conductors where the chemistry of single
atoms is extremely important in determining the current-voltage
characteristics of the whole sample.~\cite{mioprl}

In this Letter we describe a field-theoretic approach to calculate quantum
shot noise in nanoscale conductors where the electronic and ionic
distributions are calculated self-consistently. In particular, we derive an
expression of shot noise in terms of single-particle wavefunctions. The latter
quantities can then be determined using a scattering approach within the
density-functional theory of many-electron systems.~\cite{langt} As an example
we apply the method to the study of shot noise in Si atomic wires between
metal electrodes. We find that shot noise is strongly nonlinear as a function
of bias and it is enhanced for very short wires due to the large contribution
from the metal electrodes, while for longer wires it shows an oscillatory
behavior for even and odd number of atoms.~\cite{tae}%

\begin{figure}
\includegraphics[width=.25\textwidth]{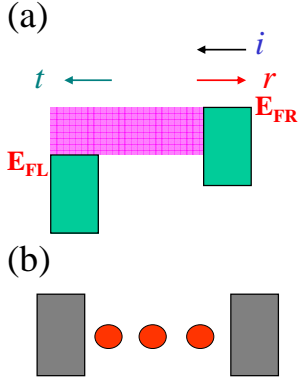}
\caption{(a) Energy diagram of the two bare electrodes kept at a certain
external potential difference. The left electrode is positively biased:
electrons incident from the right electrode are partly transmitted into the
left electrode with probability $t$, and reflected back into the right
electrode with probability $r$. (b) Schematic of the system investigated: the
sample atoms are sandwiched between two bulk electrodes.}
\label{figdemo}
\end{figure}

Let us start by writing the field operator of propagating electrons for a
sample connected to a left (L) and right (R) reservoir in terms of
single-particle wavefunctions $\Psi_{E}^{L(R)}\left(  \mathbf{r,\mathbf{K}%
_{\Vert}}\right)  $ with energy $E$ and component of the momentum parallel to
the electrode surface $\mathbf{K}_{\parallel}$~\cite{langt} (see
Fig.~\ref{figdemo} for a schematic of the system investigated)
\begin{equation}
\hat{\Psi}=\hat{\Psi}^{L}+\hat{\Psi}^{R},
\end{equation}
where
\begin{equation}
\hat{\Psi}^{L(R)}=\sum_{E}a_{E}^{L(R)}\left(  t\right)  \Psi_{E}^{L(R)}\left(
\mathbf{r},\mathbf{K}_{\Vert}\right)  .
\end{equation}
The coefficients $a_{E}^{L(R)}\left(  t\right)  =\exp(-i\omega t)a_{E}^{L(R)}$
are the annihilation operators for electrons incident from the left (right)
reservoir, satisfying the usual anticommutation relations: $%
\{a_{E}^{i},a_{E^{\prime }}^{j\dag }\}=\delta _{ij}\delta \left( E-E^{\prime
}\right) $ We assume that the electrons coming from the left (right) electrodes 
thermalize completely in the reservoirs, i.e., the statistics of electrons 
coming from the left (right) electrodes is determined by the equilibrium 
Fermi-Dirac distribution function $f_{E}^{L\left(  R\right)  }$ in the left (right)
reservoir, i.e.,
\begin{equation}
<a_{E}^{i\dag }a_{E^{\prime }}^{j}>=\delta _{ij}\delta \left( E-E^{\prime
}\right) f_{E}^{i},
\end{equation}%
with i, j=R, L.

>From the above field operator we can define the current operator
\begin{equation}
\hat{I}(z,t)=-i\int d\mathbf{R}\int d\mathbf{K}_{\Vert }\left( \hat{\Psi}%
^{\dag }\partial _{z}\hat{\Psi}-\partial _{z}\hat{\Psi}^{\dag }\hat{\Psi}%
\right) ,  \label{currentop}
\end{equation}%
where $d\mathbf{R}$ defines an element of the electrode surface. For sake
of simplicity we will assume in the following that the right Fermi level
$E_{FR}$ is higher than the left Fermi level $E_{FL}$, so that the average
value of the current at zero temperature is simply
\begin{equation}
<\hat{I}>=-i\int_{E_{FL}}^{E_{FR}}dE\int d\mathbf{R}\int d\mathbf{K}_{\Vert
}\left(  \tilde{I}_{E,E}^{R,R}\right) ,\label{average}%
\end{equation}
where
\begin{equation}
\tilde{I}_{E,E^{\prime}}^{ij}=\left(  \Psi_{E}^{i}\right)  ^{\ast}\nabla
\Psi_{E^{\prime}}^{j}-\nabla\left(  \Psi_{E}^{i}\right)  ^{\ast}%
\Psi_{E^{\prime}}^{j}\text{,} \label{cur}%
\end{equation}
with i, j=R, L.

We now calculate the shot noise as the Fourier transform of the electric
current auto-correlation function in the limit of zero frequency and zero
temperature.~\cite{buttiker} The electric current we refer to is the excess
current $\Delta\hat{I}$ with respect to the average current Eq.~(\ref{average}%
). We first consider the spectral density
\begin{equation}
2\pi S\left(  \omega\right)  =\int dte^{i\omega t}\left\langle \Delta\hat
{I}\left(  t\right)  \Delta\hat{I}\left(  0\right)  \right\rangle
,\label{spectrum}%
\end{equation}
where $\Delta\hat{I}(z,t)=\hat{I}(z,t)-\left\langle \hat{I}\right\rangle $
corresponds to the excess (with respect to the average) current operator. We
evaluate the quantum statistical expectation value in 
Eq.~(\ref{spectrum}) by using the Bloch-De Dominicis theorem~\cite{kubo} (see 
also Ref.~\onlinecite{buttiker2}) to obtain

\begin{align}
S\left(  \omega\right)   &  =\sum_{i,j=L,R}\int dEf_{E+\omega}^{i}\left(
1-f_{E}^{j}\right)  \cdot\nonumber\\
&  \int d\mathbf{R}_{1}\int d\mathbf{K}_{1}\tilde{I}_{E+\omega,E}^{ij}\int
d\mathbf{R}_{2}\int d\mathbf{K}_{2}\tilde{I}_{E,E+\omega}^{ji}\text{.}%
\label{sw}%
\end{align}
In the limit of zero frequency and zero temperature the Fermi distribution
function reduces to the step function $f_{E}^{L\left(  R\right)  }%
=\theta\left(  E_{FL(R)}-E\right)  $, and the noise power $2\pi S\left(
\omega=0\right)  =\lim_{T\rightarrow0}\int dt\left\langle \Delta\hat{I}\left(
t\right)  \Delta\hat{I}\left(  0\right)  \right\rangle $ can be written as
\begin{equation}
S=\int_{E_{FL}}^{E_{FR}}dE\left\vert \int d\mathbf{R}\int d\mathbf{K}\tilde
{I}_{E,E}^{LR}\right\vert ^{2}\text{,}\label{shot}%
\end{equation}
where the range of energy integration is from $E_{FL}$ to $E_{FR}$ and
$\tilde{I}_{E,E}^{LR}$ has been defined in Eq.~(\ref{cur}).%

Equation~(\ref{shot}) is the desired expression that relates current
fluctuations to single-particle wavefunctions. As it was previously noted, it
is clear from Eq.~(\ref{shot}) that shot noise is not simply proportional to
the conductance of the sample~\cite{buttiker} but is determined by an
interplay between electron wavefunctions incident from the left electrode and
electron wavefunctions incident from the right electrode.~\cite{precT} It is also
interesting to point out that at zero temperature and for a finite bias, there
are no electrons incident from the left electrode in the energy region between
left and right Fermi levels (see Fig.~\ref{figdemo}). The nonzero value of $S$
is thus a direct consequence of quantum mechanical statistics reflecting the
fact that current fluctuations at zero temperature are a purely quantum
mechanical effect.

Equation~(\ref{shot}) can be applied to study the behavior of shot noise in
any system once the single-particle wavefunctions are known. We evaluate these
wavefunctions self-consistently using the Lippmann-Schwinger
equation~\cite{langt}
\begin{gather}
\Psi_{E}^{L(R)}\left(  \mathbf{r,\mathbf{K}_{\Vert}}\right)  =\Psi
_{0,E,\mathbf{K}_{\Vert}}^{L(R)}\left(  \mathbf{r}\right)  \nonumber\\
+\int d\mathbf{r}_{1}\int d\mathbf{r}_{2}G\left(  \mathbf{r},\mathbf{r}%
_{1}\right)  V\left(  \mathbf{r}_{1},\mathbf{r}_{2}\right)  \Psi_{E}%
^{L(R)}\left(  \mathbf{r}_{2}\mathbf{,\mathbf{K}_{\Vert}}\right)
\text{,}\label{lipp}%
\end{gather}
where $\Psi_{0,E,\mathbf{K}_{\Vert}}^{L(R)}\left(  \mathbf{r}\right)
=\exp\left(  i\mathbf{K}_{\Vert}\cdot\mathbf{R}\right)  u_{E\mathbf{K}_{\Vert
}}^{L(R)}(z)$ are the wavefunctions corresponding to propagating electrons
incident from the left (right) electrode in the absence of the scattering
potential $V\left(  \mathbf{r}_{1},\mathbf{r}_{2}\right)  $, $G$ is the
corresponding Green's function, and $\mathbf{R}$ and z are the coordinates
parallel and perpendicular to the electrode surface, respectively.~\cite{prec}
Deep into the electrodes ($z\rightarrow\pm\infty$), the wavefunctions
$u_{E\mathbf{K}_{\Vert}}^{L(R)}(z)$ assume different scattering boundary
conditions according to their energy and parallel momentum.~\cite{langt} 
The potential $V\left(  \mathbf{r}_{1},\mathbf{r}_{2}\right) $ describes the
difference in potential between the complete system and the bare
electrodes.~\cite{langt} It is the sum of the
nuclear, Hartree and exchange-correlation potentials.~\cite{langt} We 
choose to represent the latter 
in the local-density approximation to density-functional theory.~\cite{dft}
Once the single-particle wavefunctions are calculated self-consistently using
Eq.~(\ref{lipp}), the electric current and shot noise can be calculated
using Eqs.~(\ref{average}) and~(\ref{shot}). Current-induced forces and,
consequently, the ionic distribution in steady state can also be determined
with the above wavefunctions.~\cite{mioprb}

\begin{figure}
\includegraphics[width=.48\textwidth]{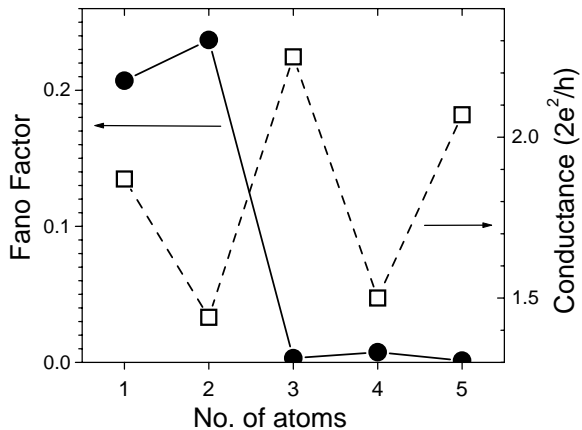}
\caption{Dependence of the Fano factor (left axis) and conductance (right
axis) on the number of Si atoms in an atomic wire between two planar
electrodes in linear response regime (bias=0.01 V).}
\label{fig2}
\end{figure}

The current operator of interest to us is the one corresponding to the extra
current due to the presence of the sample atoms between the two bare
electrodes (see Fig.~\ref{figdemo}): $\delta\hat{I}(z,t)=\hat{I}(z,t)-\hat{I}%
_{0}(z,t)$, where the first term on the RHS is defined in Eq.~(\ref{currentop}%
), and the second one is the equivalent term due to the bare electrodes only.
In the following when we discuss about the average current, Eq.~(\ref{average}),
and related shot noise, Eq.~(\ref{shot}), we refer to the extra current operator
$\delta\hat{I}(z,t)$ and corresponding fluctuations. ~\cite{langt,mioprb} 
Fluctuations of the extra current are nonlinear in terms 
of $\hat{I}$ and $\hat{I}_ {0}$ implying an 
enhancement of shot noise due to the contribution from the bare electrodes
when $\hat{I}$ and $\hat{I}_ {0}$ have comparable magnitude.


\begin{figure}
\includegraphics[width=.48\textwidth]{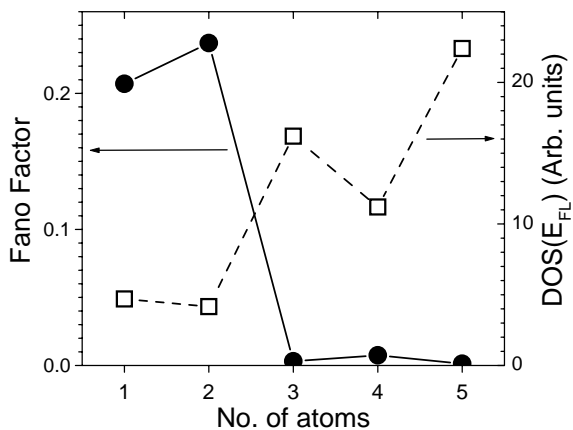}
\caption{Dependence of the Fano factor (left axis) and density of states at the
left Fermi level (right axis) on the number of Si atoms in an atomic wire
between two planar electrodes in linear response regime (bias=0.01 V).}
\label{fig3}
\end{figure}

As an example of application, we calculate the shot noise for Si atomic wires
between two bulk electrodes (see Fig.~\ref{figdemo}) as a function of the wire
length and as a function of bias. The bulk electrodes are modelled with ideal
metals (jellium model).~\cite{langt} The interior electron density of the
electrodes is taken equal to the value for metallic gold ($r_{s}\approx3$).
The single-particle wavefunctions are expanded in plane waves and convergence
of both the average current and shot noise has been checked with increasing
number of plane waves.~\cite{langt} The spacing between the silicon atoms is
fixed at $4.2$ a.u., and the distance between the electrode edge and the first
silicon atom is $2.1$ a.u., the relaxed atomic positions at zero
bias.~\cite{mioprb} We keep the ionic configuration of the system unchanged at
all different biases. 
In Fig.~\ref{fig2} we plot the Fano factor (defined as $F=S/2eI$) as a function
of the number of Si atoms for an external bias of 0.01V (linear response
regime). In the same plot (right axis) we also show the conductance in units
of the quantum of conductance $G_{0}=2e^{2}/h$. The conductance of the wires
oscillates as a function of the number of atoms.
The origin of even and odd parity oscillations is due to the fact that 
even-atom-number wires have fully occupied $\pi$ orbitals, 
while the odd-atom-number wires have a half-filled $\pi$ state at the Fermi level. 
This behavior has also been predicted for C-atom 
wires~\cite{langavou} and can be similarly explained in terms of 
lower (larger) density of states at the Fermi level of the total 
system (electrodes plus atoms) for even (odd) number of atoms.
This is shown in Fig.~\ref{fig3} where the density of states at the left Fermi level
(the right Fermi level is at 0.01eV above the left Fermi level) is plotted as
a function of the number of atoms. The Fano
factor follows a similar oscillatory pattern. It is lower
(larger) for odd (even) number of atoms, exactly opposite to the conductance
oscillations behavior. This result can be rationalized by noting that lower
conductance implies lower transmission probability and thus larger Fano factor
(see Eq.~\ref{eqbutt}). These results also indicate that current fluctuations persist 
with increasing wire length.~\cite{cherba} For the one- and two-Si atom cases, the Fano factor
is enhanced due to the fluctuations introduced by the proximity of the two
electrodes. In these two cases, the distance between the electrodes is small so that 
the current across the electrodes without the Si atoms
inbetween is of the same order of magnitude as the current of the total system 
(electrodes plus atoms). On the
other hand, the shot noise of the bare electrodes without the atoms inbetween is 
larger than the shot noise of the total system. As a consequence, the Fano factor 
has a large contribution from the two bare electrodes, and the
non-additive behavior of the total shot noise is evident in this case. 

\begin{figure}
\includegraphics[width=.48\textwidth]{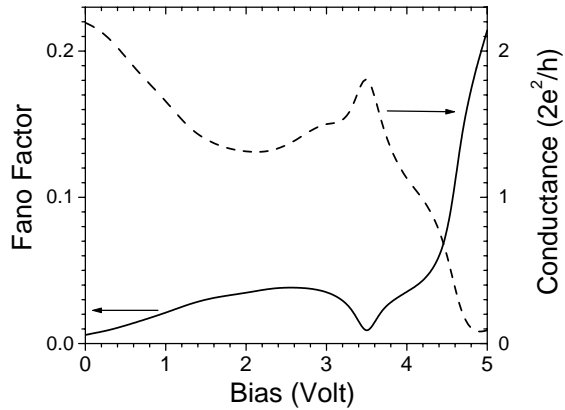}
\caption{Differential Fano factor (left axis) and differential conductance
(right axis) as a function of applied bias in nonlinear regime for an atomic
wire composed of three Si atoms placed between two electrodes as described in Fig.
\ref{figdemo}.}
\label{fig4}
\end{figure}

We now discuss the dependence of shot noise on the bias in the nonlinear regime.
We choose a 3-Si wire as an example. The differential Fano factor (defined as
$F=\partial S/\partial(2eI)$) as a function of bias is shown in
Fig.~\ref{fig4}. The differential conductance ($G=\partial I/\partial V$) is
plotted in the same figure (right axis). From Fig.~\ref{fig4} it is evident
that the differential conductance reaches a local minimum at about 2V and a
local maximum at about 3.5V. The local minimum (maximum) in the conductance is
associated to a lower (larger) density of states between the right and left
Fermi levels (not shown). The differential Fano factor follows an opposite
trend that can be explained again as due to a decreased (increased)
transmission probabilities for an electron scattering in the corresponding
energy window. The Fano factor is also strongly enhanced in the regime where the
differential conductance is close to zero. However, no complete suppression of noise is
evident from Fig.~\ref{fig4} at the bias corresponding to the conductance
peak. This is due to the contribution of several channels with different
$\mathbf{K}_{\Vert}$ that have a transmission probability lower than one at
that bias.


We would like to thank Zhongqin Yang for helpful discussions.  
We acknowledge support from the 
NSF Grant Nos. DMR-01-02277 and DMR-01-33075, and Carilion Biomedical Institute. 
Acknowledgement is also made to the Donors of The 
Petroleum Research Fund, administered by the American Chemical Society, for partial 
support of this research.


\begin{references}
\bibitem {prec}Note that the wavefunctions $\Psi_{E}^{L(R)}\left(
\mathbf{r,\mathbf{K}_{\Vert}}\right) $ have also the same labeling as the
wavefunctions $\Psi_{0,E,\mathbf{K}_{\Vert}}^{L(R)}\left(  \mathbf{r}\right) $
even though $\mathbf{K}_{\Vert}$ is no longer a good quantum number. We keep
this notation for numerical convenience.
\end{references}
\end{document}